# The Nubo Virtual Services Marketplace

James Kempf, Sambit Nayak, Remi Robert, Jim Feng, Kunal Rajan Deshmukh,
Anshu Shukla, Aleksandra Obeso Duque, Nanjangud Narendra, and Johan Sjöberg

**Abstract**—In this paper, we describe a virtual services marketplace, called Nubo, designed to connect buyers of virtual services (or tenants) with providers of those services on a cloud computing platform. The marketplace is implemented as a collection of distributed microservices along with a marketplace portal that runs as a Web application. The heart of Nubo is the Saranyu tenant and service management microservice. Saranyu is a *decentralized application* (dApp) built on top of the J.P. Morgan Quorum blockchain. Tenant and service accounts are represented as static (nonnegotiable) smart contracts written in the Solidity language. Quorum provides a tamper evident and tamper resistant distributed ledger, whereby multiple cloud and service providers can co-operate to provide service resources to tenants in a trustworthy fashion. Services offer resources consisting of a collection of attributes describing what the tenant can consume, and tenants subscribe to service resources through the Nubo Marketplace portal. The Service Manager microservice provides multitenant support for containerized services built for deployment and orchestration using Docker that were originally not designed to be managed through Saranyu. We discuss our design goals for Nubo, describe the overall architecture, discuss some details on how Saranyu uses the blockchain and smart contracts, and provide comprehensive performance and scalability data measured on the Saranyu REST API. The results indicate Saranyu is competitive with published results for comparable operations on the Havana release of OpenStack Keystone, but Saranyu provides a much richer collection of tenant and service management functionality than Keystone.

**Index Terms**—cloud computing, service delivery platform, blockchain, smart contracts, Solidity

---◆---

## 1 INTRODUCTION

PUBLIC cloud computing platforms support sophisticated tenant and service management systems coupled with service deployment and charging, whereby service providers can offer their services to third party customers and be compensated for them. These systems are unique to the public cloud platform on which they are implemented, and a deployment on one public cloud provider is not portable to another. They are also monolithic and opaque, so dispute resolution is dependent on the goodwill of the cloud provider, whereas service provisioning in regulated markets such as health care requires that the public cloud provider implement specialized procedures. In all cases, the public cloud provider acts as a trust intermediary between the service provider and the tenant.

In contrast, the open source identity management system OpenStack Keystone [1] has an open REST API that manages tenant interactions with the compute/networking/storage services provided by OpenStack. However, the APIs have no provision for third party service offerings. Usage tracking, charging, and billing are handled by separate and proprietary systems. While Keystone does provide for federated identity management [2], configuring and administering a federated cloud is complex, and how dispute resolution would work is unclear. For federated OpenStack cloud deployments, there is no clear trust intermediary, as a dispute may involve multiple cloud providers.

In this paper, we describe the Nubo[1] virtual services marketplace. The idea behind Nubo is to provide a trustworthy marketplace on which various service providers, including cloud service providers, can co-operatively offer their services in a distributed fashion, even if they compete on the service offerings themselves. The platform should not require any organization to act as a trust intermediary, but rather trust should derive from the platform itself. Removing the need for trusted intermediaries should simplify dispute resolution and regulatory actions and reduce costs. These properties naturally led to using a blockchain as the basis for the tenant and service management system behind the platform.

In the next section, we discuss our design goals for Nubo. Section 3 describes the overall architecture and provides some details on the platform components other than tenant and service management. Section 4 drills down on the Saranyu[2] [3] tenant and service manager in detail and includes a subsection on performance and scalability. In Section 5, we review previous work while Section 6 summarizes the paper and outlines possible future work.

## 2 DESIGN GOALS

The original motivation for Nubo was to provide tenant and service management for a next generation cloud compute/networking/storage management system implemented based on single system image (SSI) [4][5][6] principles for cloud. In our SSI for cloud architecture, rather than

---

- *James Kempf is at Equinix, 1188 E. Arques Ave., Sunnyvale, CA, 94085, USA but the work was completed while he was at Ericsson.*
- *Kunal Rajan Deshmukh is at Ericsson, 2755 Augustine Drive, Santa Clara, CA, 95054, USA.*
- *Sambit Nayak, Anshu Shukla, and Nanjangud Narendra are at Ericsson, CITRINE, WTC-04, BAGMANE WTC SEZ, Bangalore, Karnataka, 560048, India.*
- *Remi Robert, Jim Feng, Aleksandra Obeso Duque, and Johan Sjöberg are at Ericsson, Torshamnsgatan 23, Stockholm, 164 83, Sweden.*
- *Email addresses are jkempf@equinix.com or either <first>.<last>@ericsson.com or <first>.<middle>.<last>@ericsson.com.*

[1] "Nubo" is the Esperanto word for "cloud".

[2] Saranyu is the name of the Hindu goddess of the clouds.

a centralized management plane, cloud management is provided by a collection of distributed agents managing resources on individual Linux servers to present an abstraction to the developer of the data center as a single machine. Using a blockchain and smart contracts rather than a centralized database to provide tenant and service management seemed a natural match with the SSI for cloud architecture, since the blockchain itself is inherently distributed. The initial implementation of the compute service based on an SSI for cloud architecture, called Nefele[3] [7], is one of the first services offered through the Nubo platform. However, in the process of working on the design, we realized that the trust properties of blockchain were an even more powerful motivation for using blockchain as the basis of a cloud tenant and service management system.

The standard procedure for public cloud service providers is to offer a specific collection of cloud services from which the tenant chooses and then pays as they use the periodically audit transactions. Blockchains can be permission-less (i.e., public) and per-missioned (i.e., private). The advantage of the latter, especially for users, is that it brings in an additional layer of access control ensuring that only authorized parties can view the transactions stored in the blockchain.

We therefore pivoted toward a design that could also support VM and containerized service deployments on existing cloud platforms, in addition to the SSI for cloud platform. We added two additional services deployed in the OpenStack cluster of the Ericsson Research Datacenter (ERDC) along with the distributed compute service Nefele - a serverless function service that uses the fast DAL key-value store [9] for communication and the Berkeley RiseLab Ray [10] high-performance distributed execution framework for implementing large-scale machine learning applications.

The design goals for Nubo, both for basic cloud com-service. Many cloud providers also offer extensive "free tier" services for which the tenant pays nothing but instead has a limit or quota of how much of the service they can use per month, before they need to pay. Terms and conditions are fixed and non-negotiable. In a smart contract it is possible for these terms and conditions to be modified based on negotiations between a user and a provider. Smart contracts can in principle digitally facilitate, verify, or enforce the negotiation or performance of a contract between the user and the provider, by making the terms and conditions of the contract more visible to both the user and provider.

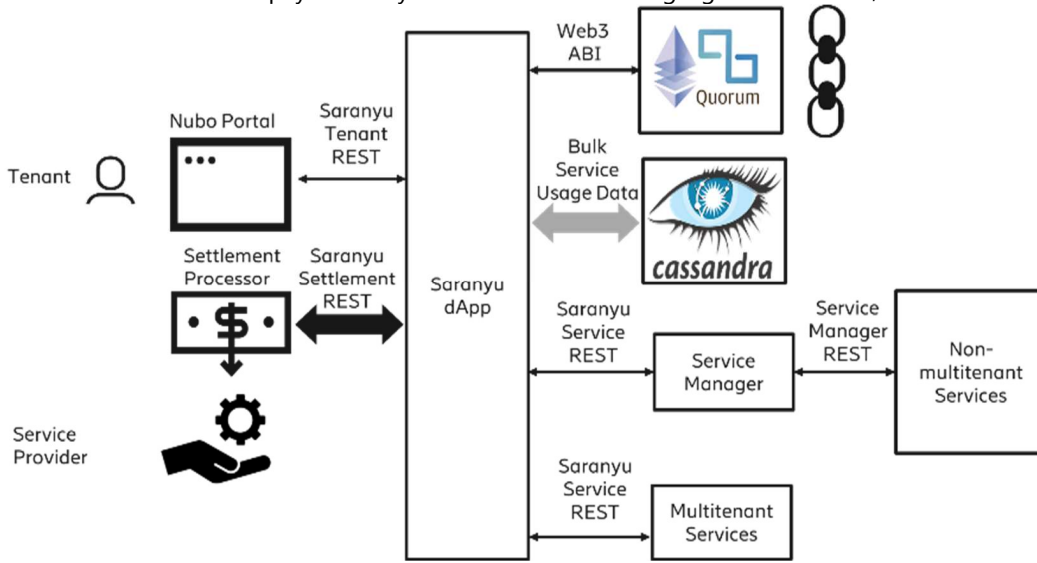

Fig. 1. Overview of the Nubo reference architecture.

In addition, blockchain technology provides a way of allowing parties in financial and other transactions to trust each other without involving any intermediaries because transactions are tamper-evident and tamper-resistant [8], therefore providing non-repudiation. While intermediaries can facilitate transactions if everything runs smoothly, they add cost and can be problematic in disputes or in regulated markets where a regulatory authority may need to

pute/network/storage services as well as for third party services built on top, are the following:

- Handle tenant and service identity and authentication when a tenant logs into Nubo or a service boots up a new instance.
- Allow services to advertise resource offerings to tenants, and for tenants to subscribe to the offerings.
- Provide tenants with proof of authorization for the services, so tenants can only utilize the services to which they have subscribed, and services can record per tenant usage.
- Collect service resource usage information for tenants, and periodically perform settlement by invoking an outside settlement processor (cryptocurrency account, credit card provider, etc.).
- Provide a convenient user portal available through a Web browser by which tenants can easily get to their service subscriptions.
- Allow multiple cloud service providers and software service providers to list their infrastructure / platform /

software services in the marketplace.

Note that while the proprietary tenant and service management systems of the public cloud providers all fulfill the first 5 goals, they do not fulfill the 6th. Also, OpenStack Keystone handles 1 and 3, but only for basic compute/networking/storage services and not for third party services.

### 3.1 Nubo Marketplace Portal

The Nubo Marketplace portal is a Javascript application written using the Angular toolkit implementation of Ericsson's Brand 2.0 UI look and feel[12]. A tenant establishes an account and logs in through a login page where a public key is used to identify the tenant to Saranyu. The tenant must also provide access to a private key that is used to sign service subscription contracts and messages to

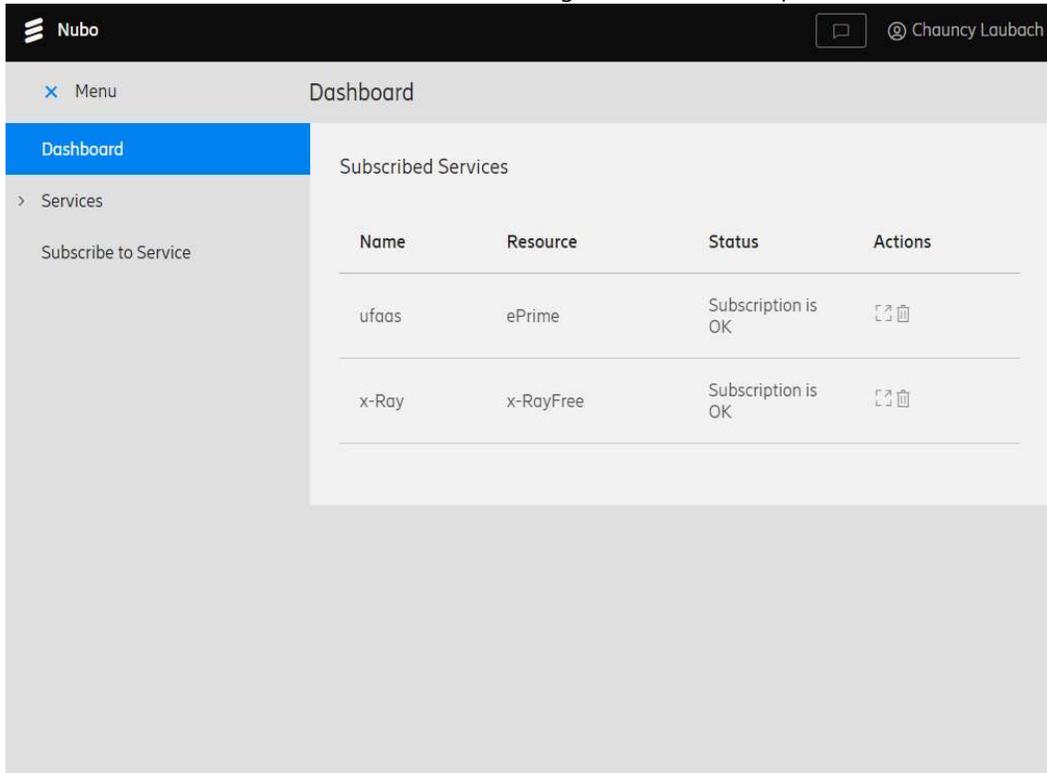

Fig. 2. Example Nubo summarized overview dashboard.

## 3  NUBO REFERENCE ARCHITECTURE

The Nubo reference architecture is shown in Fig. 1 and consists of the following four parts:

- The Nubo portal, a Web application that runs in a browser and provides the tenant/user with a convenient interface to the services registered with the Saranyu tenant and service management system,
- The Saranyu dApp that implements the tenant and service management REST API as a microservice and communicates with the Quorum blockchain,
- The Quorum blockchain microservice that records the tenant and service smart contracts, the service resource offerings, and the service contracts for tenant subscriptions,
- The Cassandra [11] distributed NoSQL database that stores bulk service usage data for reporting and billing,
- The Service Manager, a microservice which handles the Saranyu service REST API for services that were not designed to be multitenant.

In the following subsections we briefly describe the Nubo Marketplace portal, the Service Manager, and Quorum. Saranyu is described in Section 4 in more detail.

Saranyu; however, the private key never leaves the user's device. The portal consists of three basic panels:

- A summarized overview dashboard panel, an example of which is shown in Fig. 2, where the tenant's subscriptions are listed and from which the tenant can access the service through a service URL or delete their subscription,
- A detailed per service dashboard panel where the tenant can view the attribute usage and charges, if any, for the service,
- A panel where the tenant can subscribe to new services through a three-step process.

In addition, an experimental version of Nubo features a billing panel where a tenant can look at their current and past bills.

To subscribe to a service, the tenant first clicks on the new service panel menu item, and a list of services comes up to which the tenant can subscribe. The tenant then clicks on a service for a display of the service resources. A service can offer multiple tiers of service as multiple resources, like a free tier with limited attribute consumption, or different levels of paid tiers where the tenant needs to pay for attribute consumption. The tenant selects a resource, confirms the selection, and the Nubo UI calls through to the Saranyu

Tenant REST API to generate a subscription. If the subscription completes successfully, the tenant can then access the service through the service URL.

## 3.2 Service Manager

Many services were not originally implemented to be multitenant. If the service is packaged using Docker containers, for isolation, and deployed and orchestrated using Docker, then the Service Manager can provide multitenant support and integration with Saranyu with minimal modification to the service code.

Fig. 3 provides a more detailed view of the Service Manager architecture. Each type of service is deployed in a separate VM with a Service Manager instance handling the mul-connection, it takes the callback on the service's behalf, and must arrange for the service's resource manager to be notified. The service URL is how the tenant accesses the service from the Nubo portal. More details on the callback and service URLs can be found in Section 4.3.

Saranyu implements a "push" model for usage data collection, where the service must push its usage data to Saranyu over a REST call. For services that were not developed with Saranyu in mind, the Service Manager instead implements a "pull" model, where usage data is collected from wherever the service happens to make it available. Some services may include code for fine grained metrics collection and recording to a key-value store (KVS) or database, while for others, usage collection may depend on Linux cgroups or other statistics.

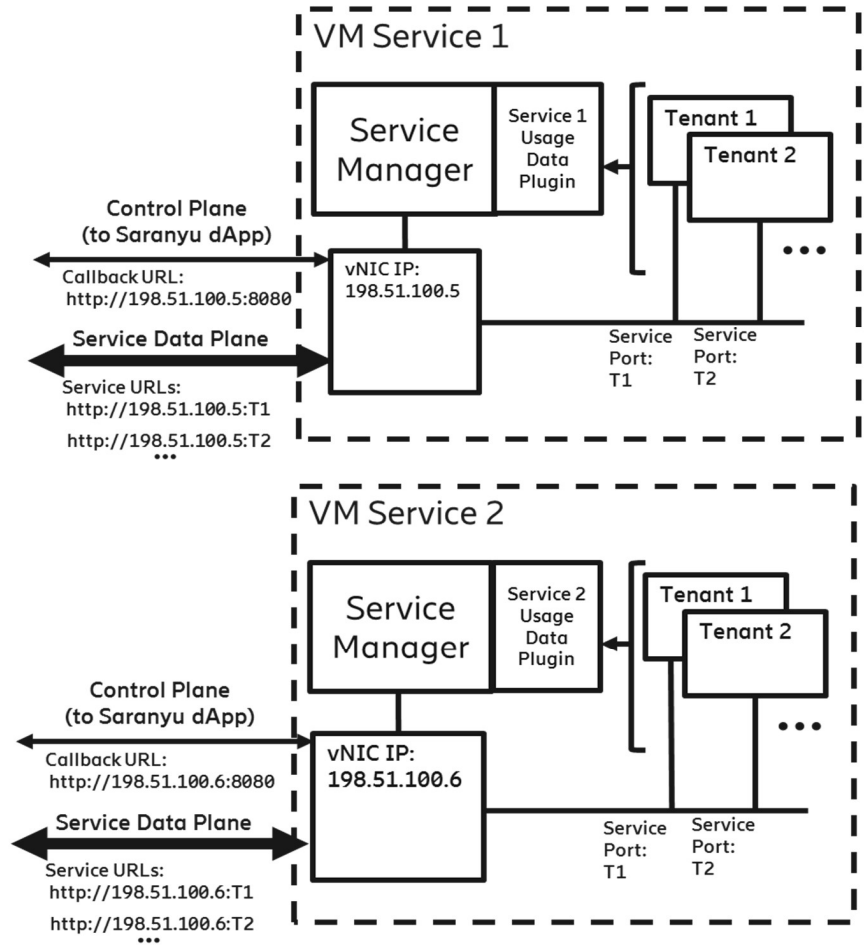

Fig. 3. Details of the Service Manager architecture.

titenant control plane connection back to/from Saranyu and the callback URL from Saranyu. The Service Manager takes care of creating the service's account in Saranyu, and registering the service resource offering, which requires the Service Manager to handle the service's public/private key pair for registering in Saranyu, in order to sign the service offering. The service instances are deployed in separate Docker containers, one for each tenant, and they handle the service URLs for the tenants on separate ports. In general, Saranyu notifies the service when a new tenant subscribes to or unsubscribes from a service in order that the service can manage resources. If the Service Manager is handling the Saranyu control plane

The Service Manager handles this with a plug-in architecture, as shown in Fig. 3.

## 3.3 Quorum

Quorum [15] is an open source port of the public permissionless Ethereum distributed ledger system to a permissioned setting, where nodes need to be whitelisted to join the Quorum network. As such, Quorum inherits the order-execute architecture of Ethereum, where transactions are first ordered and broadcast to each node then executed on the nodes. However, because only whitelisted nodes can join the network, the distributed consensus protocol need not be partially

Byzantine fault tolerant but only crash fault tolerant, a somewhat weaker criterion that does not cover traitors but is much faster and requires a minimum network size of only three nodes rather than four as is the case for partially Byzantine fault tolerance. Quorum features pluggable consensus, and Saranyu uses the RAFT algorithm [16] which is crash fault tolerant. Smart contract execution is handled by the Ethereum Virtual Machine (EVM) running in the Go-ethereum (geth) service, exactly like in the public Ethereum network, except no cryptocurrency (gas) is required to execute smart contract code.

Just like the Ethereum public network, Quorum has two kinds of accounts: external accounts and contract accounts. External accounts belong to some external entity like a person or a company, are controlled by a public key, and can trigger transactions. Contract accounts have code associated with them (smart contracts), and the code can be triggered by calls from other contracts or from an external account. The code can execute arbitrarily complex operations but can only read and write state from and to the blockchain.

Quorum has been in production use since 2016 and is quite stable, which is why we chose it as the basis for Saranyu. Recent benchmarking has shown that Quorum can be tuned to achieve over 1600 tps [19] which is sufficient for our application, as the performance data in Section 4.6 should demonstrate.

## 4 SARANYU TENANT AND SERVICE MANAGER

As shown in Fig. 1, Saranyu is the heart of Nubo and provides tenant and service management for the marketplace. The Saranyu Javascript dApp code handles communication through REST APIs with the tenants, services, and settlement processors handling different sorts of payment credentials while on the backend, the Web3 SDK [13] calls into Solidity smart contract objects in Quorum.

### 4.1 Functional Architecture

In overall operation, Saranyu works as shown in Fig. 5. In the current prototype, only one external account exists on the Quorum blockchain, corresponding to the Saranyu dApp for a single cloud provider. Tenants and services hold contract accounts that contain smart contracts enforcing the terms of their accounts.

In Step 1, tenants establish accounts in Saranyu through the Nubo Marketplace portal, providing the following basic identifying information: the tenant name, an email address and mobile phone number for contact, a charging credential which is simulated in the prototype by an Ethereum account address on a privately deployed Ethereum blockchain, and the public key part of a public/private key pair generated with the ECC256 algorithm used by Ethereum [20]. Tenant accounts are only identified by their public keys. Tenants authenticate with Saranyu through the Nubo UI using their public key and a message signed with their private key, and are issued a time limited JSON Web Token (JWT) [21] which is self-validating, indicating their right to subscribe to services and access the services to which they have subscribed.

Likewise in Step 1, services also establish accounts by registering the following through the Saranyu REST API: the service name; a JSON document containing resource offerings to which tenants can subscribe in a format described in Section 4.3; a callback URL where Saranyu can post a notification when a new tenant subscribes or an existing tenant unsubscribes; a service URL which the Nubo Marketplace portal can display for tenants to click through on when they want to access the service; a settlement account identifier where tenant charges for resource usage can be paid, likewise simulated with the private chain Ethereum account number; and a public key. Service accounts are also only identified by their public keys. The service resources are delegated to Saranyu using the grant-delegation model described in Section 4.4, so Saranyu can sub-delegate them to tenants.

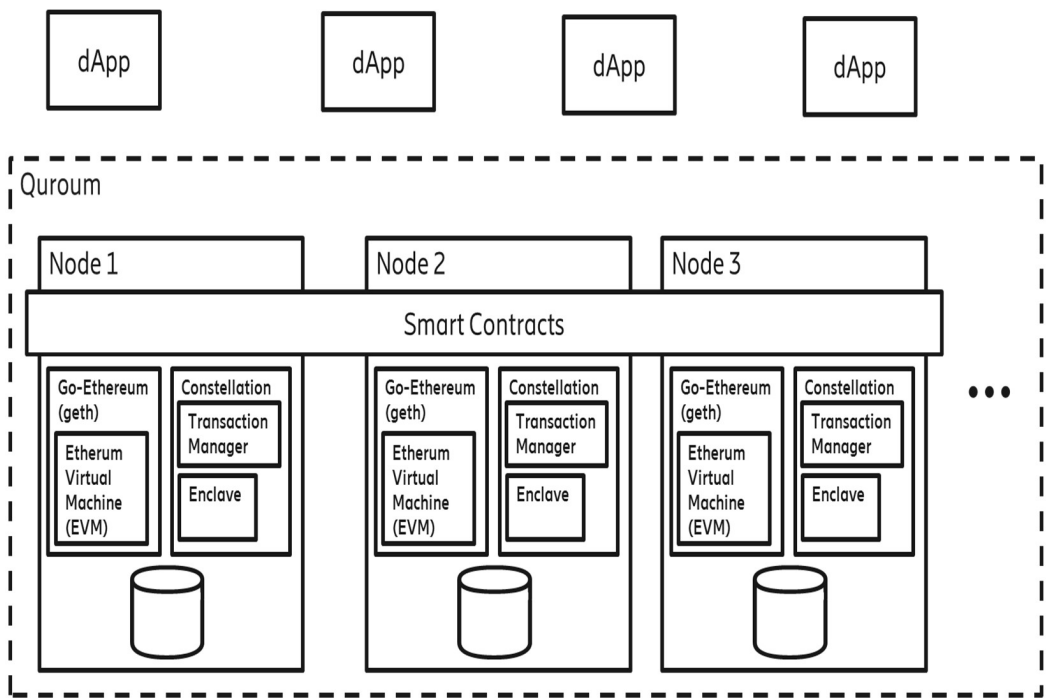

Fig. 4. Quorum architecture including Constellation private blockchain [14].

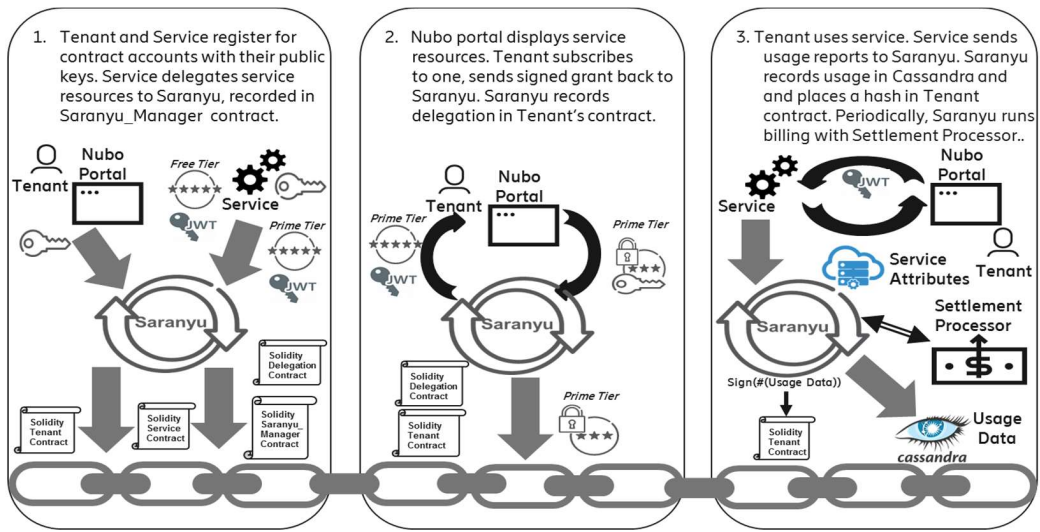

Fig. 5. Saranyu functional architecture, showing tenant, service, and billing actions.

In Step 2, the Nubo portal displays the services and their resources available to the tenant by querying Saranyu. The tenant selects a service resource and clicks through the subscription process to subscribe. When a new tenant subscribes to a service (or cancels an existing subscription by unsubscribing), the service receives a callback on its service URL including the resource to which the tenant has subscribed. It is then up to the service to control tenant resource usage to ensure that any quota is enforced. The service can reject the subscription if insufficient resources are available to fulfill it. Saranyu does not insert itself into the service API calls or resource usage tracking to avoid a costly outcall to a network service in the middle of what could be a performance sensitive resource allocation operation.

In Step 3, the tenant clicks through from the Nubo portal on the service URL to access the service. The Nubo portal includes the tenant's authorization JWT in the parameters to the service URL. A service must check the JWT's validity, and if there is any concern about the possibility of a token being stolen, then the service must require the tenant to use HTTPS and check that the signature on the HTTPS message and on the token are from the same key. Additionally, services requiring any further authorization must arrange for that themselves. For example, if a service has different types of access for different classes of tenants, like employees with different roles in an organization, then the service URL should link to a service-specific login page where the tenant needs to provide additional credentials. As the tenant uses the service, it accumulates charges for resource attributes that the service has marked as chargeable (called *metrics*) and runs up quota on attributes which are quota limited. The service reports usage on a per tenant basis to Saranyu, and Saranyu stores usage records into the Cassandra distributed NoSQL database system [11]. A signed hash of the records is deposited in the blockchain, to ensure tamper-proofness and non-repudiation[4]. Tenants can obtain a service specific dashboard view of their resource usage in the Nubo Marketplace portal.

Services include an attribute that specifies a charging interval in their resource description documents, which also covers metrics. Payment in the current prototype is handled manually, the tenant must transfer cryptocurrency from their wallet on the private Ethereum blockchain to the service's wallet. Saranyu listens for the payment operation on Ethereum (by registering an event handler for the payment event) and marks the bill as paid. A tenant can also obtain a list of current and past bills through the Nubo Marketplace portal. The design accommodates alternative payment options (credit card billing for example) though none are currently implemented [3].

## 4.2 Saranyu REST API

The Saranyu REST API consists of three generic groupings of calls:

- Calls used by services to register, deregister, and report tenant resource usage, labelled "Saranyu Service REST" in Fig. 1,
- Calls used by the Nubo portal on behalf of the tenant to create and delete a tenant account, to subscribe or unsubscribe to a service, to query for service resource usage, and to fetch a bill, labelled "Saranyu Tenant REST" in Fig. 1,
- Calls involving billing, mainly by the Nubo portal to register a payment. In the current prototype, there are no explicit calls to settlement processors because the experimental cryptocurrency billing implementation requires the tenant to take care of that by hand, but this may be automated in the future. This API is labelled "Saranyu Settlement REST" in Fig. 1.

The following subsections have high level descriptions of these calls.

---

[4] We originally stored the usage data directly in Quorum but the performance of bulk data storage directly into the blockchain was insufficient.

### 4.2.1 Saranyu Service REST API

The Saranyu service REST API consists of the following calls:

- `/services POST` – register a new service with Saranyu. Parameters include a public key for the service; a signed JWT with the array of resource descriptions (see next section); and a JWT with the callback URL where Saranyu can call to tell the service about a new registration or unregistration, the service URL, which contains a link to the service's Web application, and the service's Ethereum account for payment. Creates a new service smart contract object for the service and puts it on the list of active services.
- `/services/{id} DELETE` – delete a service registration. The service can either include its public key in the end-point designation or as a parameter. Deletes the service's smart contract object from the list of active services.
- `/authentication POST` – Return a JWT allowing the service to access Saranyu. Parameters include the service public key, a string identifying that the authentication is for a service, and a JWT with a *not-too-skewed* timestamp signed by the service to validate the public key identity.
- `/usage/metrics POST` – send a record of chargeable resource attributes (metrics) to Saranyu for the usage period. Parameters include the service's public key and a signed JWT for the metrics record.

```
resource-definition =
"resource" ":" "{"
    "name" ":" string ","
    "simple_attributes" ":" "[" attr-def-list "]" ","
    "renewable_quota_attributes" ":" "[" quota-attr-def-list "]" ","
    "nonrenewable_quota_attributes" ":" "[" quota-attr-def-list "]" ","
    "metrics" ":" "[" metric-def-list "]" ","
    "usage_tracking_interval" ":" number "," # in seconds
    "charging_interval" ":" number # in seconds
"}"

attr-def-list = attr-def attr-def*

attr-def =
"{" "name" ":" string ","
    "value" ":" string ","
    "description" ":" string
"}"

quota-attr-def-list = quota-attr-def quota-attr-def*

quota-attr-def =
"{" "name" ":" string ","
    "unit" ":" string ","
    "quota" ":" number ","
    "description" ":" string
"}"

metric-def-list = metric-def metric-def*

metric-def =
"{" "name" ":" string ","
    "unit" ":" string ","
    "rate" ":" number ","
    "currency" ":" three-letter-currency-symbol ","
    "description" ":" string
"}"

#internationally recognized currency symbol, like USD, EUR, BTC, etc.

three-letter-currency-symbol = string

string = #collection of Unicode or ASCII characters
```

Fig. 6. BNF for JSON resource description

- `/usage/quota POST` – send a record of quota limited resource attributes to Saranyu for the usage period. Parameters include the service's public key and a JWT for the quota record.

### 4.2.2 Saranyu Tenant REST API

The Saranyu tenant REST API consists of the following calls:

- `/tenants POST` – Create a new smart contract object for the tenant and register on the list of active tenants. Parameters include the tenant's public key, used as the account identifier, contact information (name, phone number, email address), and payment credential (currently only an Ethereum account number is allowed).
- `/tenants/{id} DELETE` – Remove the tenant's smart contract object from the list of active tenants. The tenant's public key is used to identify the account and can be either appended to the endpoint or included as a parameter.
- `/tenants/{id} GET` - Return the information for the tenant identified by either the id appended to the end-point name or included as a parameter.
- `/authentication POST` – Return a JWT allowing the Nubo portal to access Saranyu on behalf of the tenant. Parameters include the tenant public key, a string identifying that the authentication is for a tenant, and a JWT with a *not-too-skewed* timestamp signed by the tenant to validate the public key identity.
- `/grants POST` – Request a grant of service attributes. Parameters include the service public key, the tenant's public key, a string containing the name of the resource type, and a requested duration and number of allowed subdelegations (see Section 4.4 for more information on the grant-delegation resource consumption model). The return includes a JWT signed by Saranyu with the grant, along with a JSON document describing the grant.
- `/delegations POST` – Request a delegation of service attributes. Parameters include the grantor (typically Saranyu's) public key, the JWT signed by the grantor returned by the `/grants` call, the tenant public key, a JWT signed by the tenant with the JSON document for the grant, a string containing the name of the resource type, and a requested duration and number of allowed subdelegations (see Section 4.4 for more information on the grant-delegation resource consumption model).
- `/delegations GET` – Request a list of delegations matching the public key of the tenant passed as a parameter.
- `/delegations DELETE` – Delete a resource delegation corresponding to the service, tenant, and resource name passed as parameters.
- `/usage/metrics GET` – return a metrics report for the tenant's usage of the resource over a given period. Parameters include the service's public key, the tenant's public key, the resource name, a string indicating whether a consolidated or detailed report is desired, and two optional parameters indicating the starting and ending timestamp of the period.
- `/usage/quota GET` – return a quota attributes report

for the tenant's usage of the resource. Parameters include the service's public key, the tenant's public key, and the resource name.

- `/billing/bills GET` – return list of bills matching the URL parameters. Parameters include the service public key, the tenant public key, name of the resource type, and additional parameters indicating whether the bill should be for a specific time period, or identifying the bill by an id, and whether the bill has been paid or not.

### 4.2.3 Saranyu Settlement REST API

This API has only one call in it, `/billing/payment POST`, and it is made by the Nubo portal acting as the settlement agent to register a payment verification with Saranyu. Using the current crypocurrency charging mechanism, it causes Saranyu to listen to the private Ethereum blockchain for a deposit to the service's Ethereum account. When Saranyu detects the payment, it marks the bill as paid. The parameters are the service and tenant public keys, name of the resource, id of the bill being paid, and details of the payment, i.e. whether it is in ether and if so, how much.

### 4.3 Service Resource Description

Saranyu supports a JSON service resource description allowing services to advertise their resource offerings to tenants. Fig. 6 contains a BNF for the resource description.

A resource description has two required attributes, the charging interval and the usage tracking interval. The usage tracking interval gives the time, in seconds, during which the service usage is measured. At the end of the usage tracking interval, the service sends resource usage records to Saranyu. The charging interval attribute is the time, in seconds, before Saranyu totals up the charges for the resource and calculates a bill.

A resource description can also have four types of service defined attributes:

- Simple attributes, which provide a single unstructured value. An example is a fixed 20 GB ephemeral disk size

of a VM offered by a free tier service.

- Renewable quota attributes, which are renewed after the expiration of the charging interval. If tenant usage hits the quota, the tenant needs to wait until the quota renews before using more. An example is 2000 of CPU seconds per charging period.
- Nonrenewable quota attributes, which are absolute limits on the amount of resource the tenant can use. An example is a quota of 10 VMs maximum.
- Metrics, which have a charge attached to them. Usage is metered, and the charge totaled up at the end of the charging period for billing.

### 4.4 Resource Consumption Model and Protocol

Saranyu utilizes a two-step *grant-delegation* model for resource consumption by the tenant [22]. The model requires the tenant to first ask for a grant of the resources, then sign a JWT for the resources indicating their acceptance of the terms (quota, pricing on the charged for attributes, etc.) and request a delegation of the resources from Saranyu to them. The JWT acts as a static (non-negotiable) contract between the tenant and the service, with the blockchain as the trust intermediary. Figs. 7 and 8 illustrate the protocol.

In Fig. 7, a service registers with Saranyu by issuing a `/services POST` REST command. The resource description JWTs contain the service's JSON resource description documents in the format shown in Fig. 6. Saranyu creates delegation smart contracts of Solidity type `Delegation` in Quorum, delegating the resources from the service to Saranyu for each of the resources that the service offers. The delegation contracts contain the JWTs provided by the service in the `/services POST` call. These service-to-Saranyu delegation contracts serve as the root of the delegation contract chain, for subsequent tenant delegations of the service resources, and are recorded in Saranyu's `Saranyu_Manager` contract.

In Fig. 8, when a tenant requests a grant for a service resource through the Nubo Marketplace portal, the portal issues a `/grants POST` REST command to Saranyu. Saranyu then returns the service's JSON resource description document and a signed JWT for the resource, indicating that it has authorization to issue the grant. The tenant generates a signed JWT from the JSON document through the portal, indicating acceptance of the grant, and the portal issues a `/delegations POST` REST call to request a new delegation contract on behalf of the tenant, including the grantor JWT it received from Saranyu in the `/grants` response and the grantee JWT it generated for the tenant.

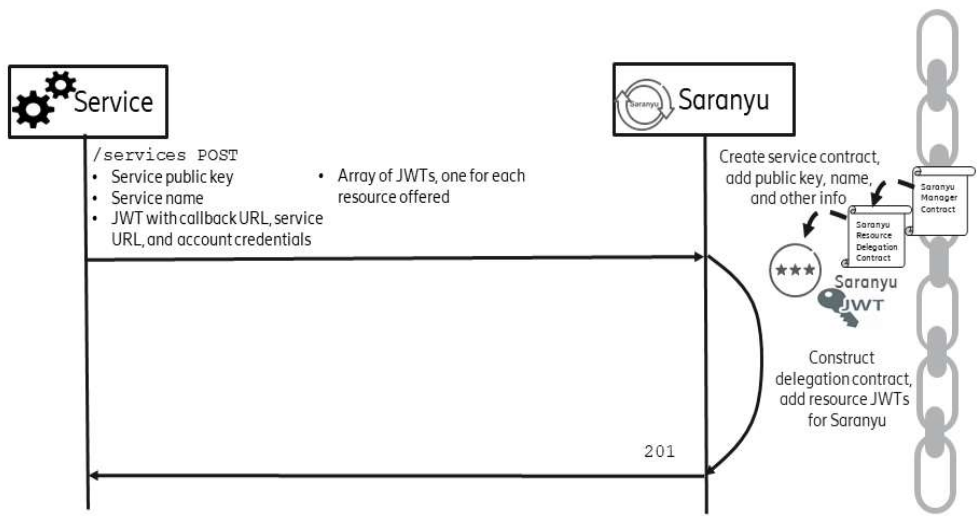

Fig 7. Service registration.

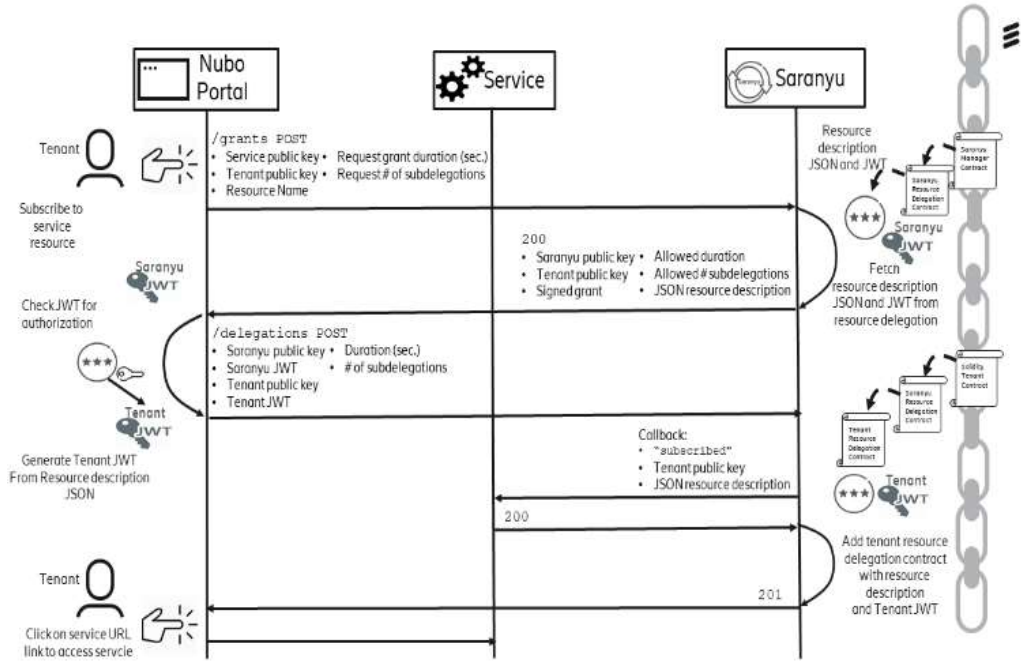

Fig. 8. Tenant subscription.

During the `/delegation` call, Saranyu issues a callback to the service, allowing the service to account for the resource usage, and returns a positive response only if the service approves the resource usage. If the callback returns successfully, Saranyu generates a delegation contract for the tenant and links it to the parent. The grant-delegation model allows traceability of resource grants back from a tenant to the original service controlling the resource, through a linked chain of Delegation Solidity contract objects and their JWT fields.

Although the current prototype only supports one level of subdelegation, the design accommodates extending the delegation chain, whereby the tenant could subdelegate the resource to another party. This would allow an enterprise tenant to request grants of resources for a collection of employees, then subdelegate the resources to the employees for example. The model also has room for delegations to be revoked or suspended if, for example, a tenant doesn't pay its bill and for delegations to time out after a time limit is exceeded, though these functions aren't implemented in the current prototype. More information on these design extensions can be found in [3].

### 4.5 Usage Tracking and Billing

Tracking of service attribute usage and periodic billing are two other important functions provided by Saranyu. Most public cloud providers don't make the technical details of their billing and usage tracking systems public but compared to an example OpenStack billing system [23], the current Saranyu prototype is much less featureful and would require additional work to make it production ready. In particular, the current prototype has fixed rates via the JSON service resource metrics description and fixed charging and billing without any scope for dynamic rating, rate negotiation, coupons or discounts. Saranyu does, however, let the service charge for and report whatever usage attributes it wants, and does not limit the service to common IaaS/PaaS usage tracking like some charging and billing systems for OpenStack do.

Fig. 9 illustrates the protocol for usage tracking. Saranyu provides a REST endpoint, `/usage`, for a service to record service resource attribute usage and for the Nubo portal to query for usage to display on a tenant's service-specific dashboard. The service reports metrics usage through the `/usage/metrics PUT` command, while the `/usage/metrics GET` command returns an array of metrics usage records to the Nubo portal. Each record has the following format:

```
[
    {
        "metric": string, #metric name
        "unit": string,
        "rate": integer,
        "currency": string,
        "usage": [
            {
                "unitsUsed": integer,
                "charge": integer,
                "start_timestamp": integer,
                "end_timestamp": integer
```

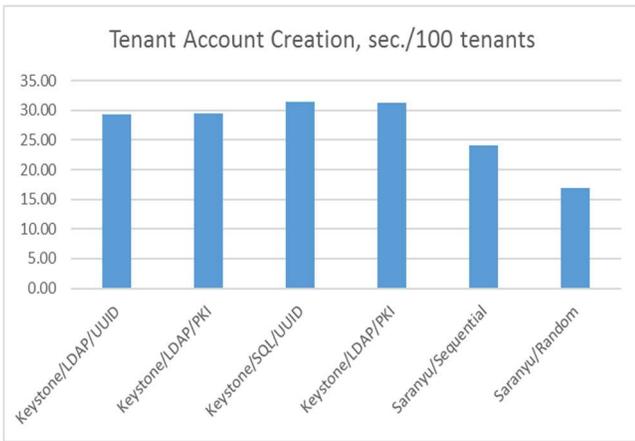

Fig. 10. Performance comparison of Saranyu and OpenStack Keystone v2, Havana release for tenant creation [25].

```
            }
          ]
        }
      ]
```

The `/usage/quota` URI provides a way for a service to report and the Nubo portal to query for service re source attributes that are quota limited, both renewable and nonrenewable. The service reports quota attribute usage with the `/usage/quota PUT` command. The `/usage/quota GET` returns an array of JSON objects for the quota attributes formatted as follows:

```
[
  {
    "quota_attribute": string, #attribute name
    "unit": string,
    "unitsUsed": integer,
    "quota": integer
  }
```

We have also implemented an experimental billing service for Saranyu based on paying the bill with Ethereum cryptocurrency, available at the `/billing` endpoint. The `/billing/bills GET` command allows the Nubo portal to query by tenant and service for unpaid bills or all bills in a particular time period. The `/billing/payment POST` command allows Saranyu to receive a callback from a listener registered on the private Ethereum blockchain when an ether transfer from the tenant account to the service account occurs, so the bill can be marked as paid. More support is needed for automatic settlement, as the current prototype requires the tenant to make the payment manually through some Ethereum wallet, but this will likely require incorporating such wallet functionality into Saranyu. The Quorum blockchain also allows the retrieval of stored data for completed billing transactions using the Web3 SDK, in particular, using APIs such as `getTransaction(txHash),` enabling cloud service operators to learn about past usage and billing history of tenants.

### 4.6 Performance and Scalability

Saranyu performance and scalability were tested on a four VM deployment running on the OpenStack Rocky release in the Ericsson Research Datacenter (ERDC). The VM instances were running Ubuntu 18.04 with 8 vCPUs, 16 GB of memory, and 20GB of ephemeral disk. The VMs were all run with anti-affinity constraints, to force scheduling on different servers. Blockchain data was written to the VMs' ephemeral disks.

Three of the VMs were running Saranyu, Quorum Maker v2.5.1 [24] and Quorum v2.1.1. Quorum Maker allows easy management of a multi-node Quorum network from a single Web portal. The consensus protocol run with Quorum was RAFT [16]. The fourth VM ran the benchmark load generator. Three benchmark runs were performed, and the measurements reported here are the average of the multiple runs (3 for the Havana comparison and 2 for the scalability study). Time was measured on the client side in the benchmark VM before the requests were sent and after all outstanding requests completed.

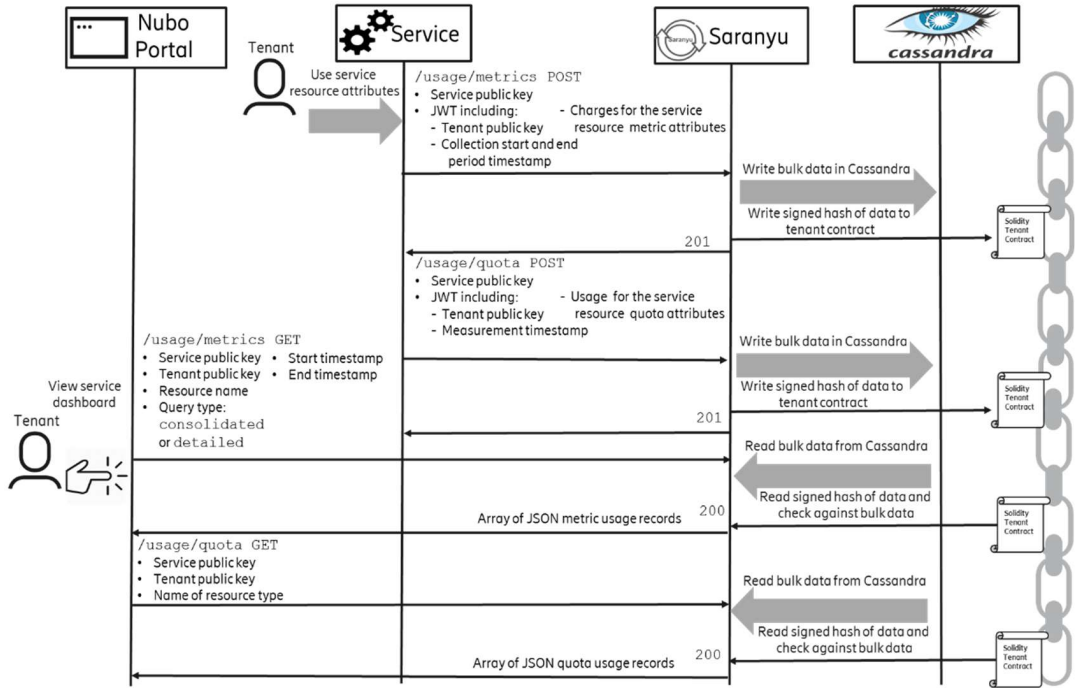

Fig. 9. Metric and quota resource attribute usage tracking.

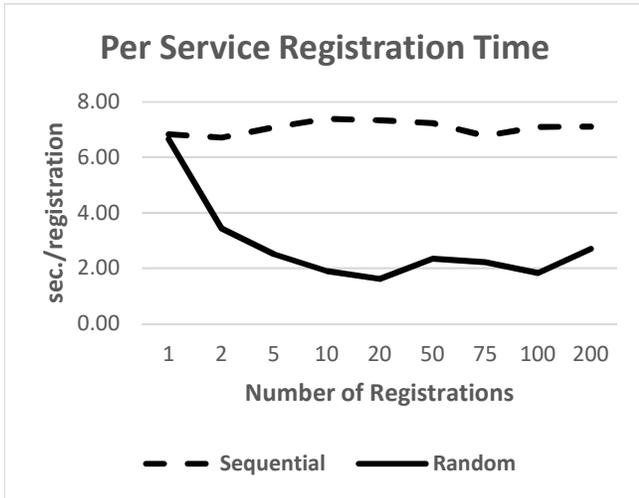

Fig. 12. Scalability of service registration.

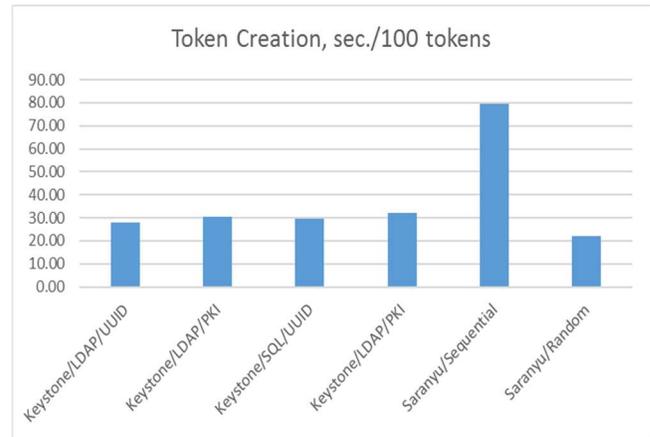

Fig. 11. Performance comparison of Saranyu and OpenStack Keystone v2, Havana release for token creation [25].

The tests were run with two different load generation regimes:

- Sequential—the load generator sent the requests sequentially to a single Saranyu node, waiting for the return of the REST call before sending another, resulting in one REST operation per transaction block.

- Random—The load generator sent each of the requests in a run in parallel to one of the three Saranyu nodes randomly chosen for each request, then waited for all requests in a run to complete; and the time for all the requests to complete in parallel was measured. Such load regime may result in the batching of multiple transactions in a transaction block.

Note that the sequential regime is similar to how centralized database system without replication would work and reflects the measurements made on OpenStack Havana Keystone from the literature [25] so it is a fairer comparison, but the random regime is similar to how Saranyu would be used in a production deployment.

### 4.6.1 Comparison with OpenStack Havana

For comparisons of tenant creation and authentication, we searched the literature on OpenStack Keystone and found one performance study of Keystone v2 on the Havana release[5] of OpenStack [25] that included tenant creation. The study was run on Ubuntu 13.10 with most services disabled, except for Keystone, MySQL, and LDAP. Although the author doesn't say if Ubuntu was run directly on the hardware, other performance tests reported in the same study were run directly on the hardware. In comparison, the study reported here on Saranyu was run on VMs, as described above.

The Havana study measured the time taken for tenant creation, token generation, and token validation.

Measurements were made for Keystone running on two different backend database systems, LDAP and SQL, and two different token generation/validation schemes, UUID and PKI. Since Saranyu uses JWTs, which are self-validating, we report the Saranyu results here for tenant creation (`/tenants POST`), which registers a new tenant by creating a record for the tenant in the blockchain, and authentication (`/authentication POST`), which creates an authorizing JWT.

Fig. 10 shows the results for the tenant creation operation (`/tenants POST` REST call). Saranyu tenant creation time is less than Keystone v2 for the sequential regime and is about half that of the Keystone v2 reported time for the random load generation regime, how a production deployment would function. In Fig. 11, the results for token creation (`/authentication POST` REST call) are shown. Saranyu is about 2.8x slower than Keystone v2 for the sequential regime, and slightly faster for the random regime, again how a production deployment would function.

Overall, Saranyu compares favorably on these two operations with an early OpenStack release in a measurement regime that best mimics a production deployment. If a similar amount of optimization were to be performed on Saranyu as has been performed on OpenStack over the last five years, performance is likely to improve substantially.

### 4.6.2 Scalability Study

---

[5] Havana was released in 2013/2014.

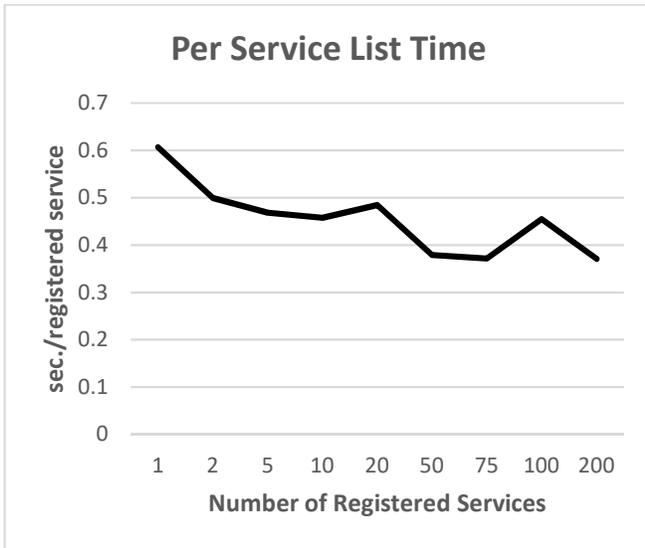

Fig. 14. Scalability of service list.

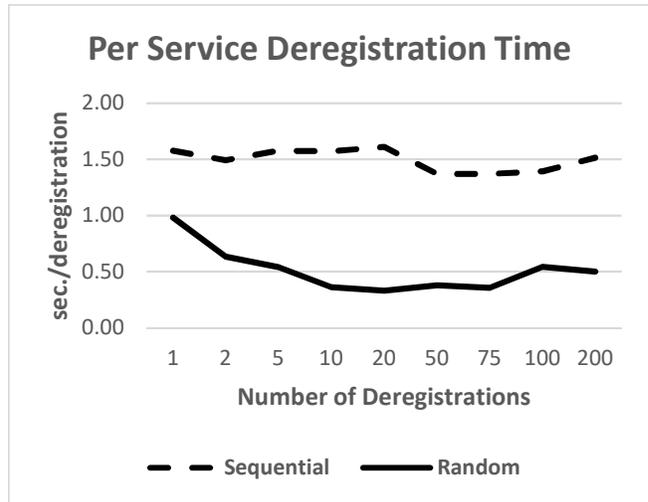

Fig. 13. Scalability of service deregistration.

Scalability was measured by measuring the amount of time taken for the following operations:

- Registering a service (`/services POST`) and deregistering a service (`/services DELETE`) as a function of the number of calls,
- Listing services (`/services GET`) as a function of the number of registered services,
- Subscribing to a service resource by obtaining a rights grant delegation (`/grants POST`, `/delegations POST`) as a function of the number of tenants subscribing.

Each operation was run for 1, 2, 5, 10, 20, 50, 75, 100, and 200 (or also additionally 150 and 300 in some cases) instances of the measured variable of interest, i.e. calls for registration and deregistration, and number of registered services for list services, and tenants subscribing in the case of subscriptions. Two timing measurements were made, the average of the two calculated for average total

Two service resources were used for `/services POST` and `/services DELETE`. One resource, `small-process`, had two simple attributes, one renewable quota attribute, one non-renewable quota attribute, and two metric attributes. The other resource, `large-process`, had two simple attributes and two metric attributes. The resource used in the `/grants` operation was `small-process`.

In Figs. 12 and 13, the results are shown for the `/services POST` and `/services DELETE` calls. The time per operation is roughly constant for the sequential regime. For the random regime, the time decreases as the number of services increases.

In Fig. 14, the results of the `/services GET` measurement are shown. Unlike the other measurements, for the services list operation, the load generator sent the requests in parallel to a single node rather than selecting a node at random. Again, the time per service decreases as the number of registered services increases.

Fig. 15 contains the results of the subscriptions scalability study. The graphs show the total time required to subscribe to a service by obtaining a rights grant delegation as the sum of the per tenant averages for the `/grants POST` and the `/delegations POST` operations. The same trends as in the previous measurements can be seen. The sequential time is constant and the random time decreases as the number of simultaneous tenant subscriptions increases. The per tenant subscription time for the random regime, is around 400 ms. at minimum.

Overall, the scalability results for Saranyu look quite promising. None of the measured operations shows an exponential increase in time as the number of operations or services increases. For the random regime and for list services, the time actually decreases as the number of operations increase, probably due to amortization of the blockchain transaction overhead, though at some point it will reach a minimum. Naturally, for a large cloud system in a production environment with many thousands of services, additional work would be needed to ensure performance was maintained.

## 5 RELATED WORK

Saranyu's rights delegation architecture was inspired by the Berkeley WAVE IoT rights delegation work [22]. WAVE is an IoT identity management, authentication, and authorization service defined on the public Ethereum blockchain for allowing rights delegation of access to IoT devices. For example, the owner of an apartment who signs up with AirBnB can delegate rights to the electronic lock to AirBnB, which can then sub-delegate to customers that rent the apartment.

Another problem that has motivated our work is how to reduce hardware costs and software licenses/maintenance costs on a cloud using multi-tenancy. For example, [26] presents an architecture for achieving multi-tenancy by enabling users to run their services in a multi-tenant SOA framework as well as providing an environment to build multi-tenant applications. If services deployed in the cloud can all share the same tenant and service management system as in Saranyu, the cost of deploying a service is reduced and the service provider can pass the savings along to their customers.

There are a few systems like Nubo that support service delivery based on blockchains and smart contracts. Bletchley [27] is a project in Microsoft Azure to provide PaaS services for authentication, authorization, key issuance, storage, access, and lifecycle management, with the goal of enhancing Azure's marketplace to allow tokenized assets. Similarly, the IBM Watson IoT platform [28] provides an open source smart contract system based on Hyperledger Fabric [29] for IoT use cases. The platform also includes a data mapping component to route the data to the blockchain contract. These are system services designed for application developers to design, implement, and deploy blockchain-based applications, and not end user systems for matching service providers to customers like Nubo nor do they explicitly provide for resource usage tracking and settlement from the platform.

Finally, the big public cloud providers all have marketplace services for developers to offer services to end users and other developers and have charging and billing handled by the platform [30][31][32]. These are production systems that handle millions of transactions per day across multiple availability zones. However, because they depend on opaque databases run by the cloud provider, if a dispute or a regulatory issue arises, the tenant is dependent on the goodwill of the cloud provider. Transparency is lacking into how these services function. With Nubo, multiple parties in a cloud or other service provider consortium can run the Saranyu dApp and write to the blockchain. Because the blockchain is tamper resistant and transparent, trust can be established between the parties and between them and their customers.

## 6 CONCLUSIONS AND FUTURE WORK

In this paper, we have presented the Nubo virtual services marketplace prototype. Nubo allows tenants to find and purchase access to virtual services such as a compute service or a streaming media service, and for service providers to offer such services for compensation from their tenants. The heart of Nubo is Saranyu, a tenant and service manager based on a permissioned, distributed ledger, or private blockchain. Tenants and services sign up for accounts which are recorded on the blockchain as smart contracts. Services delegate resources to Saranyu, which then subdelegates them to tenants. Services can offer resource attributes that are quota limited or for which the tenant needs to pay, and Saranyu can handle the billing. Nubo includes support for integrating both services that natively support multitenancy and services that don't, through the Service Manager. Performance measurement indicates that Saranyu is competitive with the performance of OpenStack Keystone from the Havana release, an early version of OpenStack that was at approximately the same stage of development as Saranyu currently.

One possible criticism of this work is that the blockchain is really only being used as a database, and since the Nubo marketplace in the prototype is being run by a single entity, namely the cloud provider, one could just as well use a database with a Solidity programming interface and achieve the same results. This argument ignores one of the most important properties of the blockchain, that the records are tamper-evident and tamper-resistant [8]. These properties mean that in case of a dispute, the blockchain can be audited by a third party to determine what actually happened. Regulators, auditors, and mediators can be given permission for read-only access to the blockchain in order to monitor and audit activity. Industries which are heavily regulated, like financial services and health care, especially benefit, which is one reason smart contracts and blockchains are being investigated heavily by the financial services industry.

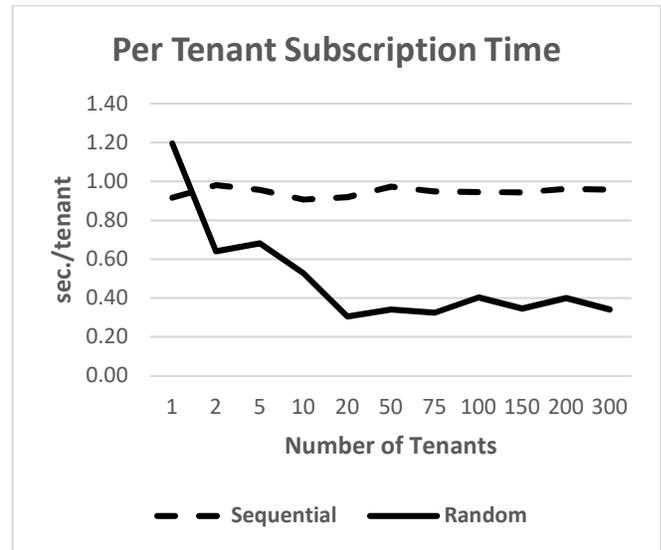

Fig. 15. Scalability of subscription.

In addition, while the Saranyu dApp only runs on a single provider in the current prototype, there is no architectural reason why multiple providers could not run the dApp in a consortium. Blockchains are particularly well suited to cases where multiple parties form a business ecosystem in which they cooperate in some cases and compete in others [8], and can establish trust between participants in the business ecosystem and their customers. Saranyu could be extended to support such an ecosystem by allowing other entities than a single cloud provider to run the Saranyu dApp on the blockchain.

Future work in Nubo would be to implement a Web portal for service providers in addition to the current portal for tenants, that would automate the service on-boarding process. The service provider could describe, register a service, and on-board their service through a Web application, by providing for example, the container images for the service, or for already registered services, check the list of current tenants. For the Service Manager, future work could include implementing more sophisticated kinds of orchestration systems, like Kubernetes [33] or orchestration into a public cloud provider. For Saranyu, future work could include completing the charging system so multiple types of credentials can be accepted, and settlement can proceed automatically without any tenant intervention unless a payment is rejected by the settlement processor. This could involve using the Quorum private blockchain Constellation for storing charging credentials since they need to be kept private. Extending Saranyu to multiple parties as mentioned above, implementing sub delegation, and implementing selective authorization so enterprise tenants can selectively authorize certain services to employees in particular roles are additional features that could be added.

Expanding Nubo out from the current focus on retail-oriented, unnegotiated contracts, would also be part of future work. In particular, many of the basic blockchain/smart contract properties discussed in Section 2, such as negotiated smart contracts and mechanisms to enable dispute resolution and regulatory monitoring should be added. Any additional work in this direction should include implementing support for the new European GDPR directive on privacy [34]. Currently, Saranyu stores contact information on the blockchain which interferes with the tenant's right to be forgotten. Finally, even though Saranyu will benefit from continuing improvement in blockchain technology performance, additional performance improvements in Saranyu need to be explored if Nubo is to be used in a production setting. In particular, we will be scaling up Nubo to handle larger data sets.

## ACKNOWLEDGMENT

The authors would like to thank Pontus Sköldström, Joacim Halén, Daniel Turull, Mina Sedaghat, Torgny Holmberg, Zoltan Turanyi, and other members of the Ericsson Research Cloud 3.0 team for their help in getting the first set of services integrated with Nubo.

## REFERENCES

[1] OpenStack Foundation, "OpenStack Docs: Keystone, the OpenStack Identity Service". [Online]: https://docs.openstack.org/keystone/latest (Accessed 2018-11-14).

[2] OpenStack Foundation, "OpenStack Docs: Federated keystone". [Online]: https://docs.openstack.org/security-guide/identity/federated-keystone.html (Accessed 2018-11-14).

[3] S. Nayak, N. C. Narendra, A. Shukla, and J. Kempf. "*Saranyu*: Using Smart Contracts and Blockchain for Cloud Tenant Management", *Proceedings of the 2018 IEEE 11th International Conference on Cloud Computing (CLOUD)*, pp. 857-861, IEEE, 2018.

[4] R. Buyya, T. Cortes, and H. Jin, "Single System Image", *Int. J. High Perform. Comput. Appl.*, vol. 15, pp. 124–135, May 2001.

[5] M. Schwarzkopf, M. P. Grosvenor, and S. Hand, "New Wine in Old Skins: The Case for Distributed Operating Systems in the Data Center", in *Proceedings of APSys*, 2013.

[6] P. D. Healy, T. Lynn, E. Barrett, and J. P. Morrison, "Single system image: A survey", *J. Parallel Distrib. Comput.*, vol. 90-91, pp. 35–51, 2016.

[7] W. John, et al. "Making cloud easy: design considerations and first components of a distributed operating system for cloud", *Proceedings of the 10th USENIX Conference on Hot Topics in Cloud Computing*, USENIX Association, 2018.

[8] D. Yaga, P. Mell, N. Roby, and K. Scarfone, "Blockchain Technology Overview", National Institute of Standards and Technology Internal Report 8202, October 2018.

[9] G. Németh, D. Géhberger, and P. Mátray, "DAL: a locality-optimizing distributed shared memory system", *Proceedings of the 9th USENIX HotCloud*, 2017.

[10] U.C. Berkeley, "Ray – RISE Lab". [Online]: https://rise.cs.berkeley.edu/projects/ray (Accessed 2018-11-14).

[11] A. Lakshman, and P. Malik, "Cassandra: a decentralized structured storage system", *ACM SIGOPS Operating Systems Review, 44*(2), pp.35-40, 2010.

[12] Stockholm Design Lab, "Ericsson – Stockholm Design Lab". [Online]: http://www.stockholmdesignlab.se/ericsson/ (Accessed 2018-11-14).

[13] Web3, "web3.eth.abi – web3.js 1.0.0 documentation". [Online]: https://web3js.readthedocs.io/en/1.0/web3-eth-abi.html (Accessed 2018-11-14).

[14] J.P. Morgan Chase, "quorum-docs/Quorum_Architecture". [Online]: https://github.com/jpmorganchase/quorum-docs/blob/master/Quorum_Architecture_20171016.pdf (Accessed 2018-11-16).

[15] J.P. Morgan Chase, "Quorum". [Online]: https://www.jpmorgan.com/country/US/en/Quorum (Accessed 2018-11-17).

[16] D. Ongaro and J. Ousterhout, "In search of an understandable consensus algorithm", *Proceedings of the USENIX Annual Technical Conference (ATC)*, pages 305–320, 2014.

[17] J.P. Morgan Chase, "Github - jpmorganchase/constellation: Peer-to-peer encrypted message exchange". [Online]: https://github.com/jpmorganchase/constellation (Accessed 2018-12-04).

[18] Phan Son Tu, "Data structure in Ethereum | Episode 3: Patricia trie". [Online]: https://medium.com/coinmonks/data-structure-in-ethereum-episode-3-patricia-trie-b7b0ccddd32f (Accessed 2018-12-05).

[19] A. Baliga, I. Subhod, P. Kamat, and S. Chatterjee. "Performance Evaluation of the Quorum Blockchain Platform." *arXiv preprint arXiv:1809.03421*, 2018.

[20] G. Wood, "Ethereum: A Secure Decentralised Generalised Transaction Ledger", *EIP-150*, Etherum Foundation. [Online]: https://gavwood.com/paper.pdf (Accessed 2018-11-18).

[21] Auth0.com, "JSON Web Tokens – jwt.io". [Online]: https://jwt.io/ (Accessed 2018-11-18).

[22] M.P. Andersen, J. Kolb, K. Chen, D.E. Culler, and R. Katz. "Democratizing Authority in the Built Environment", *Proceedings of BuildSys*, November 8-9, 2017, Delft, The Netherlands.

[23] Cyclops, "Cyclops Architecture". [Online]: http://icclab.github.io/cyclops/architecture.html (Accessed 2018-12-05).

[24] Synechron-finlabs, "quorum-maker: Utility to create and monitor Quorum nodes". [Online]: https://github.com/synechron-finlabs/quorum-maker (Accessed: 2018-11-22).

[25] Openstack Foundation, "Keystone – Performance". [Online]: https://wiki.openstack.org/wiki/KeystonePerformance#Performances_results (Accessed 2018-11-23).

[26] A. Azeez, S. Perera, D. Gamage, R. Linton, P. Siriwardana, D. Leelaratne, S. Weerawarana, and P. Fremantle, "Multi-tenant soa middleware for cloud computing", *Proceedings of the 3rd International Cloud Computing Conference,*, pp. 458–465, 2010.

[27] Microsoft, "Introducing Project 'Bletchley'". [Online]: https://github.com/Azure/azure-blockchain-projects/blob/master/bletchley/bletchley-whitepaper.md (Accessed 2018-11-24).

[28] IBM, "IBM Watson IoT". [Online]: https://github.com/ibm-watson-iot/ (Accessed 2018-11-24).

[29] Linux Foundation, "Hyperledger Fabric – Hyperledger". [Online]: https://www.hyperledger.org/projects/fabric (Accessed 2018-11-24).

[30] Microsoft, "Microsoft Azure Marketplace". [Online]: https://azuremarketplace.microsoft.com/en-us/marketplace/ (Accessed 2018-11-24).

[31] Amazon, "AWS Marketplace: Homepage". [Online]: https://aws.amazon.com/marketplace (Accessed 2018-11-24).

[32] Google Cloud, "Google Cloud Platform Marketplace Solutions". [Onlsie]: https://cloud.google.com/marketplace/ (Accessed 2018-11-24).

[33] Cloud Native Computing Foundation, "Production Grade Container Orchestration – Kubernetes". [Online]: https://kubernetes.io/ (Accessed 2018-11-24).

[34] European Union, "EUGDPR-Information Portal". [Online]: https://eugdpr.org/ (Accessed 2018-12-09).

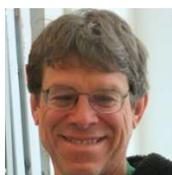

**James Kempf** graduated with a PhD in Systems and Industrial Engineering, Computer Science minor, from University of Arizona in 1984 and promptly went to work in Silicon Valley. He has worked for 24 years at a number of Silicon Valley companies prior to his current position as a Principal Researcher at Ericsson Research in Santa Clara. Dr. Kempf worked for 10 years in IETF, was a member of the Internet Architecture Board for 2 years and chaired 3 working groups that developed standards for the mobile Internet. He is the holder of 26 patents, author of 3 books, and of many technical papers. Dr. Kempf is a Senior Member of IEEE since 2018.

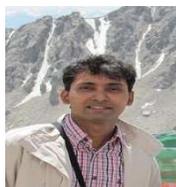

**Sambit Nayak** holds a B.E. (Hons.) Computer Science from the Birla Institute of Technology & Science, Pilani, India. He has 12+ years of professional experience, having worked at Ericsson, Oracle and Sun Microsystems in the areas of high-availability clusters and disaster recovery, systems software, and cloud infrastructure technologies. Since 2017, he is with Ericsson Research, Bangalore, India, focusing on cloud data management and blockchain technologies.

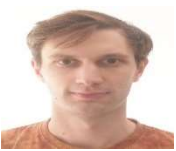

**Remi Robert** received the Engineer Degree (2015) in electrical engineering from the École supérieure d'électricité, Paris, France and Master's (2015) in electrical engineering from the Royal Institute of Technology (KTH), Stockholm, Sweden. In 2016, he joined Ericsson Research, Sweden, where he is working on service delivery and distributed cloud.

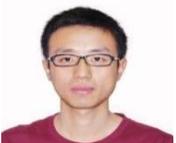

**Jim (Jinhua) Feng** received the master's degree in Network Security from Beihang University, Beijing, China, in 2008. Since 2014, he has been a researcher in Ericsson Research, Kista, Sweden. His research interests include distributed cloud and service delivery.

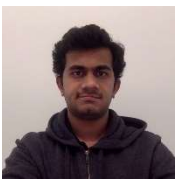

**Kunal Deshmukh** received his BE in Computer Engineering in 2014 from University of Mumbai, India. He is currently studying BE in Computer Science at San Jose State University, USA and working as an intern at Ericsson Research since June 2018. His research interests are computer vision, deep learning and blockchain.

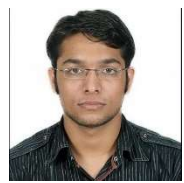

**Anshu Shukla** Anshu Shukla is a Researcher at Ericsson Research, Bangalore, India. He has a master's degree in Distributed Systems from the Computational and Data Science department (CDS) at the Indian Institute of Science (IISc), Bangalore. His current research explores abstractions, algorithms, and applications on distributed systems, spanning Cloud computing and software architectures for large-scale applications. He is also an IEEE and ACM member.

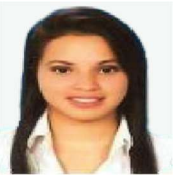

**Aleksandra Obeso Duque** received a bachelor's degree in Electronics Engineering from Universidad del Valle, Colombia, and a master's degree in Computer Science from Uppsala University, Sweden. Currently she works as a researcher in the department of Cloud Systems and Platforms at Ericsson. Her research interests include distributed systems, efficient resource management and machine learning.

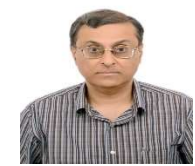

**Nanjangud Narendra** is a Principal Engineer in Ericsson Research Bangalore. Prior to joining Ericsson, he worked in Cognizant Technologies, IBM Research, HP India and Motorola India. He has about 25 years R&D experience in the Indian IT industry. His research interests span Software Engineering, Workflow Management, Web Services, Service Oriented Architecture, Cloud Computing and Internet of Things. He has published over 100 papers in international conferences and journals, is a member of the Editorial Board of Service Oriented Computing and Applications journal and a Senior Member of IEEE and ACM.

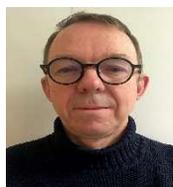

**Johan Sjöberg** received his master's degree in Electrical Engineering in 1990 and Lic. Eng. degree in Information Theory in 1994 from Chalmers University of Technology in Gothenburg Sweden. He joined Ericsson in 1993 and has mainly worked in research on speech coding, media transport and mobile communication applications. He has also worked with innovation management. He is currently in the research area Cloud Systems and Platforms.